# Enabling Effective Exoplanet / Planetary Collaborative Science
## A White Paper for the Planetary Science and Astrobiology Decadal Survey 2023-2032

**Mark S. Marley** (mark.s.marley@nasa.gov  650-604-0805 NASA Ames Research Center (ARC), Moffett Field CA, USA)

**Co-authors:** Chester 'Sonny' Harman (NASA Ames Research Center); Heidi B. Hammel (AURA) Paul K. Byrne (NCSU); Jonathan Fortney (University of California, Santa Cruz); Alberto Accomazzi (Center for Astrophysics | Harvard & Smithsonian); Sarah E. Moran (JHU Earth & Planetary Sciences); M. J. Way (NASA Goddard Institute for Space Studies); Jessie L. Christiansen (NASA Exoplanet Science Inst./IPAC-Caltech); Noam R. Izenberg (JHU Applied Physics Laboratory); Timothy Holt (Univ. of Southern Queensland); Sanaz Vahidinia (BAERI - NASA Ames Research Center); Erika Kohler (NASA Goddard Space Flight Center); Karalee K. Brugman (Arizona State Univ.)

**Co-signers:** Kathleen Mandt (Johns Hopkins Univ. APL); Victoria Meadows (U. Washington); Sarah Horst (JHU); Edwin Kite (University of Chicago); Dawn Gelino (NASA Exoplanet Science Institute/IPAC-Caltech); Britney Schmidt (Georgia Tech); Niki Parenteau (NASA ARC); Giada Arney (NASA GSFC); Julie Moses (SSI); Michael Line (Ariz. State U.); Kimberly Bott (UCR); Leigh N. Fletcher (U. of Leicester); Sarah Stewart (UC Davis); Tim Lichtenberg (Univ. of Oxford); Giada Arney (NASA GSFC); Channon Visscher (Space Science Institute; Dordt University); David Sing (JHU); Nancy Chanover (NMSU); Abel Méndez (Planetary Hab. Lab., UPR); Ed Rivera-Valentín (LPI); Tiffany Kataria (JPL); Chuanfei Dong (Princeton U.); Ryan MacDonald (Cornell U.); Sang-Heon Dan Shim (Ariz. State U.); Sarah Casewell (University of Leicester); Courtney Dressing (UC Berkeley); Aki Roberge (NASA GSFC); Emily Rickman (Univ. of Geneva); Joshua Lothringer (JHU); Ishan Mishra (Cornell Univ.); Ludmila Carone (Max Planck Inst. for Astronomy); Paul A. Dalba (UC Riverside); Scott M. Perl (NASA JPL); Paul B. Rimmer (U. of Cambridge/MRC-LMB); Joanna K. Barstow (The Open Univ.); Edward W. Schwieterman (UC Riverside); Ben W.P. Lew (U. Ariz.); Nestor Espinoza (STScI); Iouli Gordon (CfA | Harvard & Smithsonian); Thomas Greene (NASA ARC); Stephen Kane (UC Riverside); Kerri Cahoy (MIT); Jonathan Tennyson (UCL); Sergei Yurchenko (UCL); Ravi Kopparapu (NASA GSFC); Wendy Panero (OSU); Baptiste Journaux (U. Washington); Cayman Unterborn (ASU); Henry Worten (NRAO/U. Virginia); Jason Wright (Penn. State); Ehsan Gharib-Nezhad (NASA ARC); Denis Sergeev (U. Exeter); Nicholas Heavens (SSI); Diana Dragomir (UNM); Johanna Vos (AMNH); Kamen Todorov (U. Amsterdam); Beth Johnson (PSI); Michele Bannister (U. of Canterbury, NZ); Susan Mullally (STScI); Karan Molaverdikhani (Max Planck Inst. for Ast.); Bruce Macintosh (Stanford U.); Eliza Kempton (U. Maryland); Natasha Batalha (NASA ARC); Vivien Parmentier (Oxford U.); Yamila Miguel (Leiden Obs.); Joe Renaud (NASA GSFC); Joseph Llama (Lowell Obs.); Malena Rice (Yale U.); Kimberly Ward-Duong (STScI); Derek Buzasi (Florida Gulf Coast U.); Eric Nielsen (Stanford U.); Wladimir Lyra (NMSU); Kevin Hardegree-Ullman (NASA Exoplanet Sci. Inst./IPAC-Caltech); Rachel Fernandes (LPL); Monica Vidaurri (Howard U.); John O'Meara (Keck Obs.); Jonathan Brande (U. MD); Statia Cook (Columbia U./AMNH), Peter Gao (UCSC); Jake Taylor (Oxford); Eileen Gonzales (Cornell U.); Hannah Wakeford (U. Bristol); Jani Radebaugh (BYU); Michael Battalio (Yale U.); Andres Jordan (Universidad Adolfo Ibanez); Ramdayal Singh (SAC, ISRO); Giannina Guzman Caloca (UMD); Zafar Rustamkulov (JHU); Duncan Christie (U. Exeter); Stephen Thomson (U. Exeter); Miriam Rengel (MPI für Sonnensystemforschung); Roxana Lupu (NASA ARC); Jayesh Goyal (Cornell U.)

**Abstract**
The field of exoplanetary science has emerged over the past two decades, rising up alongside traditional solar system planetary science. Both fields focus on understanding the processes which form and sculpt planets through time, yet there has been less scientific exchange between the two communities than is ideal. This white paper explores some of the institutional and cultural barriers which impede cross-discipline collaborations and suggests solutions that would foster greater collaboration. Some solutions require structural or policy changes within NASA itself, while others are directed towards other institutions, including academic publishers, that can also facilitate greater interdisciplinarity.

**Introduction**
To those new to planetary or exoplanetary science it often comes as a surprise that, in practice, these are parallel fields, each with their own meetings, journals, perspectives, and even terminology. There are in effect two communities, existing side by side with only modest interaction. This white paper explores the roots of the systemic barriers that inhibit greater collaboration between planetary and exoplanetary science and suggests concrete actions to foster greater collaboration between the fields. Our ultimate vision is for there to be a single integrated field of planetary science in which knowledge, expertise, and insight is gained from all planetary systems.

Many of the themes of modern day exoplanet science are common in solar system studies and are easily recognizable by traditional solar system planetary scientists (what we will here call "legacy" planetary science). Indeed legacy planetary scientists have contributed substantially to exoplanetary studies, and much of what we know about the atmospheres, interiors, formation and evolution, and orbital dynamics of extrasolar planets has drawn heavily from traditional planetary science. However, as exoplanetary science has grown, most of the new individuals entering the field have come from astronomy, rather than planetary science, and the ties to legacy planetary science have become less prominent. This outcome is unfortunate because synergies between exoplanetary and planetary science benefit both fields, given new perspectives on longstanding planetary science questions—including, for example, linking the geological properties of planets to their potential habitability (cf. [Unterborn and Byrne et al. 2020](#))

To date, the most interesting synergies have arisen through studies of processes that are common across planets as a class; some examples are zonal atmospheric structure, atmospheric escape, and planetary migration. Studies of the atmospheric circulation of irradiated hot Jupiters predicted a strong eastward jet transporting energy from the day to nightside. As the irradiation level is reduced, additional westward and eastward jets emerge from general circulation model simulations, showing that Jupiter's counter rotating belts and zones are in reality part of a continuum of atmospheric responses to irradiation by giant planets and not a singular expression of a particular planet (Showman et al. 2015). Atmospheric escape was long suspected to be a key process sculpting solar system planets, perhaps explaining the trace atmosphere of Ganymede and the thick atmosphere of Titan. Escape is now emerging as a process of fundamental importance for extrasolar planets as well, with deep connections to our solar system understanding (e.g., Zahnle & Catling 2017). The foundational discovery of the first hot Jupiter, 51 Peg b, led to a re-examination of theories of planetary migration and ultimately to the Nice and Grand Tack models for solar system orbital evolution (Nesvorný 2018; Raymond et al. 2018). These are only a few examples and there are unquestionably many more such synergies to plumb in the future, if the communication between these fields can be better optimized.

A more mature example of synergy between studies of a solar system object and astronomy is solar astronomy. Detailed studies of the Sun set the table for our understanding of stellar structure and evolution. The standard solar model now serves as a benchmark for models of stellar structure and evolution. Later helioseismology, the study of solar oscillations, led to asteroseismology and the consequent substantial improvements in our understanding of stellar structure. As with exoplanets and legacy planetary science, solar and stellar studies use different techniques and have distinct, yet overlapping, scientific communities that may serve as a model for future collaboration between solar system and exoplanetary scientists.

While synergies such as these have contributed to the depth of exoplanetary science so far, we fear that continued separation of exoplanet science from planetary science will inhibit additional insights in the future, ultimately to the detriment of both communities. By nurturing the exoplanet–planetary science connection, we hope to ensure that more synergies emerge and advance both fields in the future. Below we briefly summarize the historical roots of the divide and offer examples of ongoing cultural differences and institutional impediments inhibiting closer collaboration. We then present a number of proposals aimed at fostering greater planetary and exoplanetary collaborations.

**The Disconnect**
*Historical Roots of the Divide*

To understand the roots of the planetary science–exoplanets divide, it helps to first briefly review the history of the fields. Aside from Earth, the first planets to be studied were of course solar system planets. The field of planetary science, ultimately rooted in the work of Tycho and Galileo and Kepler, bloomed post-World War II when sophisticated new understanding of the relevant physics and chemistry could be applied to space- and ground-based datasets made possible by new technologies. Planetary astronomers started with ground-based telescopes. Our understanding of the mass–radius diagram, planetary atmosphere thermal structure and spectra, and photochemistry, for example, all flourished from the late 1960s onward. At around the same time, the first robotic spacecraft were dispatched to our nearest planetary neighbors resolving, for example, Venus from a point of light or a disk to a world in its own right.

With the advent of robotic space exploration, however, planetary science slowly took form as its own discipline. The creation of the Lunar and Planetary Laboratory in 1960 at the University of Arizona, separated by a street—if not a chasm—from Steward Observatory, was emblematic of the trend. Planetary science began by studying planets as full disks, but as rich datasets obtained from space missions and dedicated telescopes became available, planetary scientists moved on to study highly specialized aspects of planetary phenomena, such as atmospheric escape, impact processes, and planetary internal structure.

The first generation of exoplanets were detected via the changes they induced in astronomical observations of their hosts stars. As a result, exoplanetary science arose mostly in astronomy departments, often at institutions that lacked planetary science representation. With the discovery of transiting planets and transit spectroscopy, traditional astronomical tools and approaches could now be applied to the study of exoplanets, and suddenly the nature of planetary atmospheres, interiors, and evolution was relevant to a whole new discipline. Yet these observations were predominantly obtained by astronomers, and so new approaches were invented to deal with what was soon an onslaught of data (e.g., from the *Kepler* mission). Yet those interpreting the data were generally not familiar with the planetary science literature. Likewise, planetary scientists were generally not familiar with the astronomical datasets, and also were focused on sophisticated approaches to study solar system science questions which, for exoplanets, were beyond the reach of extant astronomical tools.

This historical disconnect in methodology and specialization, while simplified here for brevity, lies at the root of the ongoing divide between the specializations. What should ideally be a single discipline of "planetary science" is largely two different disciplines failing to recognize and capitalize on synergistic opportunities (e.g., exoplanet-relevant observations by planetary missions, or exoplanet insights into solar system relevant processes). This disconnect has led to needless reinvention of the wheel or, at worst, fruitless forays into previously-explored box canyons. In the remainder of this white paper, we highlight opportunities for greater exoplanet–planetary science collaboration and suggest specific policies and actions to help bridge the gap, thereby enhancing both planetary and exoplanetary science.

*Ongoing Cultural Disconnects*

The historical bifurcation of planetary and exoplanetary science is still apparent in the culture of the fields. By its nature, exoplanetary science generally focuses on global processes detectable from full-disk observations. Spatial inhomogeneities and smaller-scale processes are inferred from phase curves or other subtle effects. Exoplanets are often understood as members of statistical classes from which gross trends are recognized and studied. In essence, exoplanetary science is at a similar stage as legacy planetary science of 50 or 60 years ago in this regard, while today most solar system planetary science focuses on spatially and/or temporally resolved studies. Likewise, exoplanet science considers atmospheric abundances to be "precisely known" if estimates are constrained to an order of magnitude, whereas legacy planetary science typically works to much higher precision, especially where in situ data are available. Further, planetary science often entails remote sensing of surfaces and atmospheres of planets for which most of the key variables (mass, gross atmospheric composition, etc.) are well understood, whereas exoplanet science often must proceed in the study of planets for which little is known beyond radius or mass. Because the techniques and background assumptions are so different, the basic knowledge base—what "everyone knows"—can be disparate as well (e.g., what does "ice giant" really mean?).

Unfortunately, such differences can lead traditional planetary scientists to not appreciate the value of their expertise or to engage with exoplanet science. As one senior planetary scientist commented regarding declining to engage, "exoplanets is the wild west". The resulting failure to engage likewise means that rank and file exoplanet scientists are not familiar with these individual planetary scientists who may have relevant expertise to a topical research problem, and instead find themselves re-inventing approaches. By encouraging greater synergy between the fields, insights can be efficiently shared, the planetary legacy can be leveraged, and the exoplanet perspective can be applied to solar system problems.

Efforts to build the needed bridges between the communities date back to the earliest days of exoplanet science (e.g., the "From Giant Planets to Cool Stars" meeting in 1999 brought together exoplanet and solar system scientists in Flagstaff, AZ), but gained little traction. We advocate for continued interdisciplinary efforts, and there are positive signs that these are happening: a notable recent effort is the "Exoplanets in Our Backyard" meeting series which opened this year at the Lunar and Planetary Institute. The goals and findings of this effort are described in the White Paper by Arney et al..

*Institutional Impediments*

The early bifurcation in the fields was exacerbated by siloed NASA funding for exoplanet science. Exoplanet research was generally supported by programs funded from the NASA Astrophysics Division. Input from the Planetary Science Division was limited; in fact, the Planetary Atmospheres Program for a time specifically identified studies of exoplanet atmospheres as being out of scope for the program, eliminating opportunities for synergistic studies. This constraint certainly contributed to the slow uptake of planetary scientists pursuing exoplanet research. Although there is much greater synergy today in SMD funding, challenges remain. Such funding structure issues are discussed more thoroughly in the complementary white paper by Mandt et al.. In short, every opportunity should be taken to reward proposals aiming to find real synergies between proposed solar system research and exoplanetary topics.

**Methods to Foster Greater Collaboration**
*Journals*

Keeping up with the increasing quantity and complexity of the scholarly literature is only one challenge our community is facing. Researchers involved in the study of exoplanets need to navigate a scholarly information landscape that is rapidly growing, yet is often disconnected from the planetary science literature. In this section we identify a set of challenges stemming from attitudinal differences in the two communities and suggest ways in which they can be addressed.

The first issue is the fact that the two communities tend to publish in separate sets of journals, with limited knowledge sharing. For the astronomy community, a major practical barrier to the legacy planetary science literature is that many key findings are published in the journals *Icarus*, *Journal of Geophysical Research*, *Geophysical Research Letters*, and chapters in the University of Arizona Press planetary science series. These venues are all paywalled and, depending on individual university holdings, can be difficult to access. Also, until recently, planetary science did not have a culture of posting to the arXiv, so even fairly recent planetary papers may not be obtainable.

Such barriers lead to unnecessary reinvention of relevant methodologies and missed warnings of common pitfalls of certain approaches. Investments in bringing the key legacy planetary science literature to open access would, if feasible, pay further dividends on decades of solar system exploration. Extending the requirement that US Federal Employee publications be made available in PubSpace to encompass the legacy planetary literature would make the substantial fraction of papers with civil servants as authors or co-authors much more readily available, particularly if PubSpace documents were retrievable through the Astrophysics Data System (ADS). Of course this would be a substantial undertaking and the implementation details would need thoughtful review. Availability of the newest journal of the AAS, the open access *Planetary Science Journal*, should be better highlighted for the exoplanet community, particularly for cross-discipline papers. Finally the increasing adoption of the arXiv by planetary scientists should continue to be encouraged.

Exoplanet papers from U.S. scientists are predominantly published in the *Astronomical Journal* and *Astrophysical Journal*. Both of these journals typically employ a single reviewer whom editors typically recruit from the astronomical community. We recommend a coordinated effort to create a pool of planetary science reviewers, sorted by specialization. Doing so would enhance cross-field collaborations and improve the transfer of planetary expertise to exoplanet science. Since exoplanets often uniquely draw on multiple diverse knowledge bases (e.g., new methods of transit lightcurve extraction and radiative transfer of atmospheric hazes), the use of more than the traditional single referee now enlisted by the AAS journals would open up opportunities for probing the planetary science knowledge base without sacrificing rigor on exoplanet-specific observing methods. The existing "statistical methods" review for the *ApJ* and *AJ* might serve as a model here. The *Planetary Science Journal* has already implemented a two-referee review process, and we encourage the extension of this practice to exoplanet papers in other journals. Given the pressures felt within the exoplanet community for rapid publication, the *PSJ* should consider methods to accelerate review where appropriate to encourage exoplanet community submissions.

Another avenue for bridging these fields would be to encourage publication of more review papers that summarize exoplanet-relevant aspects of legacy planetary science. Deep dives into particular specialities with lessons learned and summaries of best practices would be beneficial to both planetary and exoplanetary researchers. Co-authors from both domains would help keep the focus on the most important aspects for synergies without digressions into aspects likely to be unimportant for exoplanet science, given expected progressions in observational techniques. Indeed such reviews should serve as "translations", drawing explicit connections between exoplanetary and solar system topics. An upcoming special issue of *JGR - Planets* aims for such reviews.

Enhanced cross-disciplinary efforts at the information system level can also contribute by making the body of literature more apparent. At the agency level, NASA SMD has made major investments in its Astrophysics Archives and the Planetary Data System, as well as in the ADS. We note that much of the planetary geophysical literature, some of which is highly relevant to exoplanets, is not yet indexed by the ADS. As the central organizing point used to navigate new and emerging research fields, ADS can serve as a discovery platform to cross the traditional information silos. As interdisciplinary research develops, the fields can become more organically connected and discoverable through topics, citations, and co-readership, regardless of where the original research papers were published. ADS further connects the literature with data products and code, thus increasing the discoverability of both and allowing the data to be more accessible to non-experts. Increased support and enhanced collaboration across NASA's SMD archives will provide the needed infrastructure (Kurtz & Accomazzi, 2019).

*Codes and Databases*

Both exoplanetary and legacy planetary science have common needs for certain types of laboratory and theoretical fundamental data. These data include, for instance, equations of state for planet-forming materials, optical constants for clouds and hazes, molecular and atomic gas opacities, chemical reaction rates for atmospheric species, solubilities, and other geochemical relationships for planets that do not have an Earth-like composition. In some cases, there are high-quality databases appropriate for Earth interior and atmospheric studies, as well as other datasets more focused towards solar system applications. Exoplanet needs, however, often extend into physical and chemical domains not found in the solar system and thus are not well served by available datasets (see the White Paper by Kohler et al. for descriptions of experimental facilities needed to produce these data). Many exoplanet-relevant databases are becoming available, for example for molecular opacities (Tennyson & Yurchenko 2018), but such databases are generally distinct from planetary and terrestrial ones. Disjoint datasets inevitably lead to competing data formats, unnecessary complexities, and potentially erroneous results when using solar system planets as model testbeds.

Unified cross-discipline databases, curated by domain experts familiar with the needs of Earth, solar system, and exoplanetary scientists are needed to smooth collaborations and enable cross-body checks of codes. For example, it should be trivial to test an atmospheric spectral simulator against observed spectra for multiple planets with differing measured atmospheric thermal and composition profiles and different gas opacities, but the problem of disparate datasets impedes such calculations. The NASA Exoplanet Archive is now working with PDS to create a tenth "Exoplanet" node to present legacy planetary science datasets in the formats needed for direct comparison to exoplanet observations. These include phase curves as a function of incident angle for e.g. Jupiter or Venus, or transit spectroscopy of e.g. Venus or Titan. The node will also present relevant exoplanet data in formats conforming to the PDS specifications. The goal of the Exoplanets PDS node is to foster use of datasets from each domain by scientists from the other, and we encourage further investigation and development by NASA and the NSF in support of such cross-disciplinary databases.

Similarly, numerical tools used to address solar system and exoplanetary questions represent another area where there is substantial overlap, but there is often a disconnect. This paradox is largely driven by both the lack of discourse/familiarity between subject matter experts in their own disciplines, and barriers to entry for researchers coming in from outside a particular discipline. For example, general circulation models have long been used in simulating the climate of Earth, but they took some time to gain traction in the exoplanet community (see review by Heng & Showman 2015). These tools are generally repurposed from existing tools to operate outside the scope of their original design (e.g., most exoplanet GCMs started off as Earth GCMs). As such, the knowledge and expertise surrounding these tools and techniques must diffuse across the boundaries between disciplines, or researchers are forced to reinvent the wheel. The astrophysics [Asclepias project](...)

encourages scientific code to be made public and citable to facilitate discovery. Such efforts should be modeled for the planetary and exoplanetary communities.

To better enable cross-discipline model collaborations that use existing tools, or set out to build new cross-discipline ones, it is imperative that there be clear, consistent guidelines and shared, widely-known resources (similar to those we recommend for data). The issues surrounding online tool repositories are too numerous to concisely summarize here, but we note that NASA has already undertaken some action in response to previous National Academies reports (e.g., NASEM 2018) by encouraging new tools to be published as open-source software in ROSES-2020. However, this effort falls short of bridging disciplinary gaps because the tools generally are not easily discoverable and the standards for adequate documentation are sparse. Improving access to tools that have high-quality documentation would allow new researchers to identify potential collaborations regardless of discipline, to easily locate existing tools that can be used or modified to their needs, and to start using them without the months it normally takes to become familiar with undocumented tools.

The solar system science community has a deep reservoir of legacy tools, but there are many issues surrounding them, including a lack of documentation as well as difficulty finding funding to modernize them; other white papers will address these issues. When available, such tools can immediately enhance exoplanet science (e.g., the python open source exoplanet spectral generator code PICASO based on a legacy FORTRAN Titan and giant planet atmospheres code (McKay et al. 1989; Marley et al. 1999; Batalha et al. 2019)). We strongly support efforts such as these. We also support additional efforts like developer training, as well as additional funding to cover the overhead associated with continuing development, up-to-date documentation, and responding to requests from the community of new and existing users.

More generally, to enable inter-divisional/cross-topical research and discovery, we need to employ generalized modular systems, efficient pipelines, best practices, and continuous support for community-wide users. Scientists generally should not be expected to be software engineers. Again, such issues go beyond the planetary/exoplanetary divide, but as data and code-heavy endeavors these topical areas are highly immersed in such issues and we encourage the careful attention that they deserve.

*Missions*

Synergies between solar system and exoplanetary space missions can be challenging to realize. Historically, exoplanet-relevant observations by planetary missions have generally relied on individual instrument team investigators recognizing opportunities (e.g., *Cassini* observing Jupiter as an exoplanet from Saturn orbit) rather than originating from organized guidance. Indeed, exoplanet-relevant observations can seem less interesting to traditional planetary scientists unfamiliar with exoplanet science, if such opportunities are even recognized to begin with, so even relatively low cost observations can meet resistance. Likewise, traditional planetary scientist participation on exoplanet missions has generally been sparse, as exoplanet scientists are often not aware of the expertise among the planetary community.

To help elevate the importance of exoplanet-relevant observations by solar system spacecraft, we propose that all appropriate mission and/or instrument teams include designated "exoplanet participating scientist" position(s) selected through a competitive process. The exoplanet scientists would have responsibility for proposing and advocating for exoplanet-relevant observations by relevant science teams and be involved well before launch. Execution of such activities should also be a standard component of mission evaluations, including senior reviews for extended life operations. Likewise, NASA should develop criteria that ensure competitively selected representation of traditional planetary science expertise on upcoming exoplanet missions.

Ideally, as the collaboration between these fields matures and opportunities for synergies are more widely recognized, such formal collaborative structures will no longer be required. However, we view the establishment of such formal frameworks as necessary in the near term and urge NASA to implement such protocols.

*Additional Actions*

Collaborative workshops, at which exoplanet and planetary scientists jointly present topical research, should continue to be supported. These include workshops like Exoplanets in our Backyard and the Comparative Climates of Terrestrial Planets series (see white papers by Arney et al. and Soto et al.) as well as smaller sessions at planetary and exoplanetary meetings. The NASA NExSS program aims to bring together exoplanetary, planetary, and solar scientists in research collectives and is an important initiative which should continue to be supported and grown. In addition, broader-scope more tightly focused "state-of-the-field" workshops (e.g., "exoplanets for planetary scientists" and "planetary science for exoplanetary scientists") should be pursued. Summer research programs bringing together graduate students and postdocs, like the OWL workshop series at UCSC, should aim for diverse attendance by planetary- and exoplanetary-focused attendees.

**Conclusions**

Our ultimate vision is a single, unified field that uses a diversity of observational techniques to study planets near and far—the result of which will be a greater understanding of planets both as statistical populations and as individual objects. To achieve this vision we should pursue efforts that bridge the two fields—a divide that is not inherent to the disciplines but rather has emerged from accidents of history. Although the methodologies and perspectives of extrasolar planetary science will always be distinct from studies focused in our solar system, both sub-disciplines ultimately share the same goals. Only by working as a unified field can we reach our goal of broadly understanding the processes, origins, and evolution through time of all planets.

**References**

Batalha, N.E., Marley M.S., Lewis N.K., and Fortney, J.J. (2019) Exoplanet Reflected Light Spectroscopy with PICASO. *ApJ* **878** 70.
Heng, K. & Showman, A. (2015) Atmospheric Dynamics of Hot Exoplanets. *Ann. Rev. Earth & Planetary Sci.* **43** 509.
Kurtz, M. J. & Accomazzi, A. (2019) From Dark Energy to Exolife: Improving the Digital Information Infrastructure for Astrophysics.
Marley, M.S., & McKay, C.P., (1999) Thermal Structure of Uranus' Atmosphere, *Icarus* **138**, 268.
McKay, C.P., Pollack, J.B. and Courtin, R. (1989) The thermal structure of Titan's atmosphere. *Icarus* **80,** 23.
National Academies of Sciences, Engineering, and Medicine (2018) Open Source Software Policy Options for NASA Earth and Space Sciences. Washington, DC: The National Academies Press.
Nesvorný, D. (2018) Dynamical evolution of the early Solar System. *Ann. Rev. of Astronomy and Astrophysics* **56**, 137.
Raymond, S., Izidoro, A., & Morbidelli, A. (2018) Solar System Formation in the Context of Extra-Solar Planets. In Planetary Astrobiology (Editors: V. Meadows, G. Arney, D. Des Marais, and B. Schmidt) in press. arXiv:1812.01033
Showman, A.P., Lewis, N. and Fortney, J. (2015) Atmospheric circulation of exoplanets. *ApJ* **801,** 95.
Tennyson, J. and Yurchenko, S.N. (2018) The ExoMol Atlas of Molecular Opacities. *Atoms* **6**, 26.
Zahnle, K. & Catling, D.(2017) The Cosmic Shoreline: The Evidence that Escape Determines which Planets Have Atmospheres, and what this May Mean for Proxima Centauri B. *ApJ* **843,** 122.